\documentclass[aps,prl,superscriptaddress,preprint,floatfix]{revtex4}

\usepackage{graphicx,latexsym}

\begin{document}


\title{Superheating and solid-liquid phase coexistence in nanoparticles with non-melting surfaces}


\author{D. Schebarchov}
\affiliation{School of Chemical and Physical Sciences,
Victoria University of Wellington, Wellington 6001, New Zealand}
\author{S. C. Hendy}
\affiliation{School of Chemical and Physical Sciences,
Victoria University of Wellington, Wellington 6001, New Zealand}
\affiliation{MacDiarmid Institute for Advanced Materials
and Nanotechnology, Industrial Research Ltd, Lower Hutt 6009, New Zealand}



\date{\today}

\begin{abstract}
We present a phenomenological model of melting in nanoparticles with facets that are only partially wet by their liquid phase. We show that in this model, as the solid nanoparticle seeks to avoid coexistence with the liquid, the microcanonical melting temperature can exceed the bulk melting point, and that the onset of coexistence is a first-order transition. We show that these results are consistent with molecular dynamics simulations of aluminum nanoparticles which remain solid above the bulk melting temperature.
\end{abstract}


\maketitle



The thermodynamics of small systems is becoming increasingly relevant technologically as the size of devices enters the nanoscale \cite{Gross01}. A classic example of a phase transition in a small system is the melting of a free nanoparticle. Generally, one expects the melting temperature to decrease in proportion to the surface area to volume ratio of a particle because the surface free energy of a molten droplet is less than that of the corresponding solid particle. This scaling has now been confirmed in many instances \cite{Buffat76,BenDavid95,Baletto05}, although a number of exceptions to this general rule have appeared. For instance, as the sensitivity of free nanoparticle calorimetry has improved \cite{Haberland01}, a so-called `non-scaling' regime has been observed, where the melting temperature is seen to vary erratically with particle size \cite{Haberland98}. Further, small clusters of both lead \cite{Metois89} and gallium \cite{Breaux03,Breaux06} have been observed to melt at temperatures higher than the bulk melting temperature, $T_c$.  

Superheating has also been seen in the surface melting (SM) of bulk crystals \cite{Tosatti05}. Surface melting is generally thought to occur on metal surfaces for which $\Delta \gamma = \gamma_{sv}-\gamma_{sl}-\gamma_{lv} > 0$ (where $\gamma_{sv}$ is the solid-vapor (SV) interfacial energy, $\gamma_{sl}$ is the solid-liquid (SL) interfacial energy and $\gamma_{lv}$ is the liquid-vapor (LV) interfacial energy) i.e. below $T_c$ a liquid layer on the surface is frequently observed \cite{vanderVeen90}, the thickness of which diverges logarithmically as $T \rightarrow T_c$. However, for particularly low-energy metal surfaces with $\Delta \gamma < 0$ (such as Pb (111) \cite{Pluis87}, Al (111) \cite{Denier90} and (100) \cite{Frenken94}, and Au (111) \cite{Carnevali87}) surface melting is not seen, and often the surface can be heated above $T_c$ \cite{Tosatti95}. Such surfaces are called non-melting (NM). Indeed, this is thought to be the reason why supported Pb nanoparticles with only NM (111)-facets can be superheated several degrees above the $T_c$ \cite{Metois89}, although in this case the superheated solid cluster is presumably only metastable \cite{Tosatti95}.  

In an isolated nanoparticle where the internal relaxation time is less than the time for equilibration with the environment \cite{Lummen05}, the situation is somewhat different. Such a particle will follow a microcanonical caloric curve \cite{Hendy03}, where the transition between solid and liquid phases will generally occur via solid-liquid phase coexistence, even if the particle only has NM facets. In a small particle, the cost of forming the solid-liquid interface during phase coexistence is comparable to its total energy, and this can lead to an S-bend in the microcanonical caloric curve \cite{Lynden-Bell}, negative heat capacities \cite{Haberland01}, the avoidance of phase coexistence entirely \cite{Hendy05a} or dynamic coexistence between fully solid and fully liquid states \cite{Berry88}. Phase coexistence in clusters with NM facets will be particularly unfavorable and, while coexistence has been observed in molecular dynamics (MD) simulations of such clusters \cite{Cleveland94}, there is evidence to suggest that the transition between the solid and solid-liquid coexisting state in this case is first-order rather than continuous \cite{Hendy05a, Hendy05b}.

In this letter, we use microcanonical critical droplet (MCD) theory \cite{Nielsen94} to consider melting and solid-liquid phase coexistence in free clusters with NM surfaces. Interestingly, we find that not only can superheating of such solid clusters occur, but in this case, the superheated clusters are stable rather than metastable. We also find that the transition from solid to solid-liquid coexistence is first-order for NM surfaces in MCD theory, and that there is a minimum size below which solid-liquid phase coexistence is unstable. Both these findings are in agreement with results from previous MD simulations \cite{Hendy05a}. Finally, we conduct new MD simulations of solid-liquid coexistence in Al truncated octahedra (TO) with NM (111) and (100) facets, where we observe both superheating and a first-order transition to solid-liquid coexistence in agreement with the theory. In the MD simulations we use a glue potential developed for Al by the first-principles force-matching method \cite{Ercolessi94}. In particular this potential has been used to study melting of the Al (111) and (100) surfaces \cite{Tosatti95} so we expect it to be applicable here. 

\begin{figure}
\resizebox{\columnwidth}{!}{\includegraphics{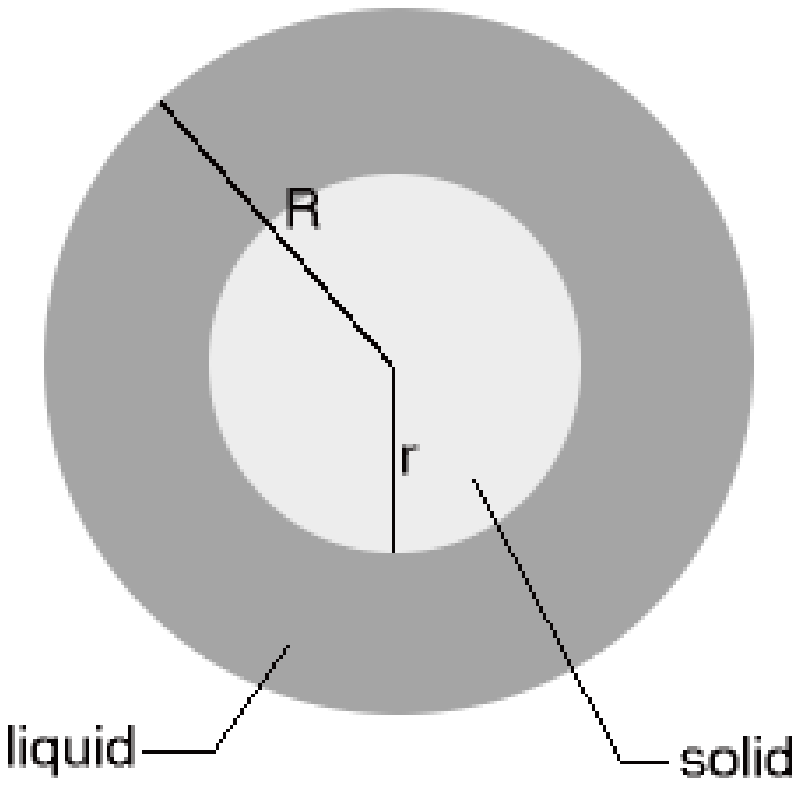}
\includegraphics{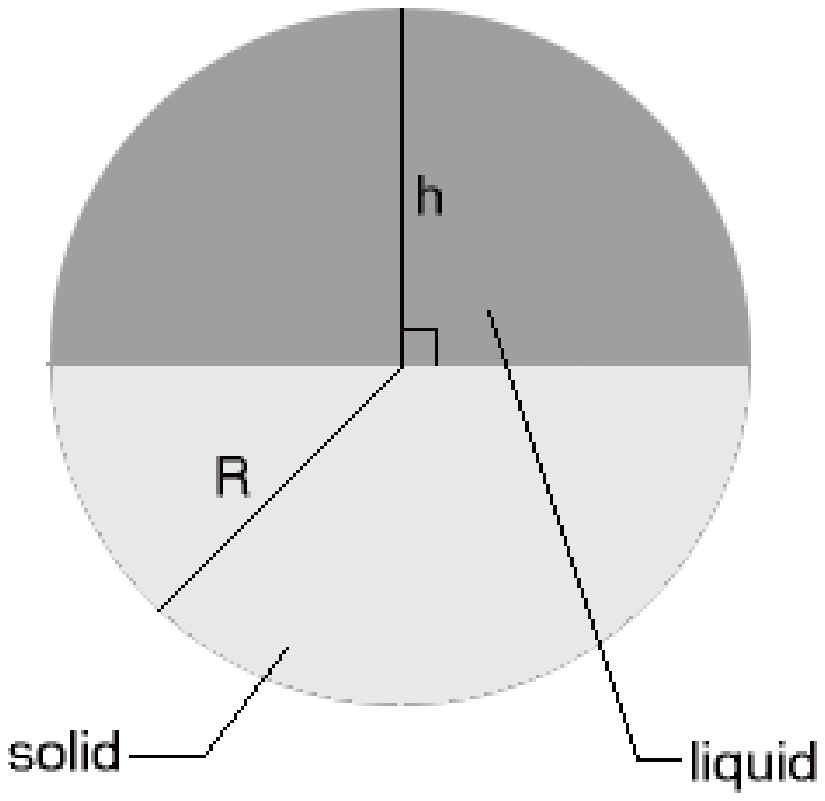}}
\caption{\label{geometry} Model geometries for coexisting clusters. For clusters with SM facets ($\Delta \gamma >0$), the spherically symmetric model on the left can be used. For clusters with NM facets ($\Delta \gamma <0$), we use the spherical cap model on the right.}
\end{figure}

For a coexisting cluster of radius $R$ with NM surface facets, the melt will not fully wet the solid so the usual assumption of a spherically-symmetric liquid layer (see Fig.~\ref{geometry}) is inappropriate. One approach is to use a geometry of two overlapping spheres \cite{Cleveland94} but such a model requires a two parameter description. Here we will use a simple one-parameter model geometry to describe the NM case (Fig~\ref{geometry}). Specifically we characterize the coexisting cluster by the height $h$ of a spherical liquid cap (we note this does resemble the solid-liquid geometry seen NM Pb nanoparticles \cite{Hendy05a}). In this model the cluster is fully solid at $h=0$ and fully liquid at $h=2R$. The total surface energy of the cluster $\Gamma$ may now be written as
\begin{equation}
\Gamma = \pi \left( 2 R \left( 2 R - h \right) \gamma_{sv} + R h \gamma_{lv} + h \left( 2 R - h \right) \gamma_{sl} \right). 
\label{Gamma}
\end{equation}
Note that we have neglected any interaction between the LV and SL interface. For short range interactions this interaction decays exponentially \cite{vanderVeen90} with a correlation length $\xi$, so $\Gamma$ will be well approximated by (\ref{Gamma}) for $h > \xi$. For Al (111), $\xi$ has been estimated to be $\sim 3$ \AA \,\cite{Tosatti95}.

Now consider the dependence of $\Gamma$ on $h$. It is easy to show that $\frac{d\Gamma(h)}{dh}<0$ only when
\begin{equation}
\label{hstar}
h > h^* = -\frac{\Delta\gamma}{\gamma_{sl}}R.
\end{equation}
Thus for a cluster with NM surfaces ($\Delta\gamma < 0$), the total surface energy can only be reduced by increasing the liquid volume for a liquid cap thickness $h > h^* >0$. This is in contrast to the SM case, where $\Gamma$ is always reduced by increasing the liquid volume. The inequality in (\ref{hstar}) strongly suggests that any coexisting state in a cluster with NM surfaces will be unstable for $h < h^*$. We note that with $\Delta \gamma < 0$, in the spherically symmetric case (Fig.~\ref{geometry}) one can show that the surface energy must increase with any increase in liquid fraction, so there is no critical liquid fraction. Nonetheless, a critical condition such as (\ref{hstar}) does not appear to be a peculiar feature of the spherical cap geometry. For instance, if one considers a solid sphere in contact at a point with a liquid droplet, one can again show there is a critical liquid fraction where surface energy will decrease with an increase in liquid fraction. Indeed the existence of a critical liquid fraction is consistent with the first-order nature of the transition to coexistence seen in simulations of Pb clusters with NM surfaces \cite{Hendy05a}.

To make the discussion more concrete, we construct a caloric curve for a cluster with NM surfaces using MCD theory  \cite{Nielsen94, Hendy05a}. The total energy $E$ of the cluster is the sum of the volumetric and surface contributions:
\begin{equation}
\label{energy}
E =  \left(V-V_l(h)\right) e_s + V_l(h) e_l + \Gamma(h)
\end{equation}
where $V=4 \pi R^3/3$, $V_l(h)= \pi/3 \, h^2 (3R-h)$ (assuming that the density is unchanged at melting) and $e_{s(l)}$ is the energy density of the solid (liquid). The total entropy, $S_m$, at $E$ is 
\begin{equation}
\label{entropy}
S_m(E,h)= \left(V - V_l(h)\right)s_s(e_s)+ V_l(h) s_l (e_l) 
\end{equation}
where $s_s(e_s)=s_s^c + c \log\left({e_s-e_s^c \over c T_c} +1 \right)$ is the entropy density of the solid region, and similarly $s_l(e_l)$ is the entropy density of the liquid. Here $c$ is the heat capacity (assumed to be the same in the solid and liquid states) and quantities with a superscript $c$ indicate values at $T_c$. At a given total energy $E$, the state of the coexisting cluster is described by the value of $h=h^\dagger$ that extremizes the entropy $S_m$. The onset of coexistence occurs at an energy when $S_m(h^\dagger,E) = S_s(E)$ and complete melting occurs when $S_m(h^\dagger,E)=S_l(E)$. These equations must be solved numerically to find the energies at which the transitions occur.  

\begin{figure}
\resizebox{\columnwidth}{!}{\includegraphics{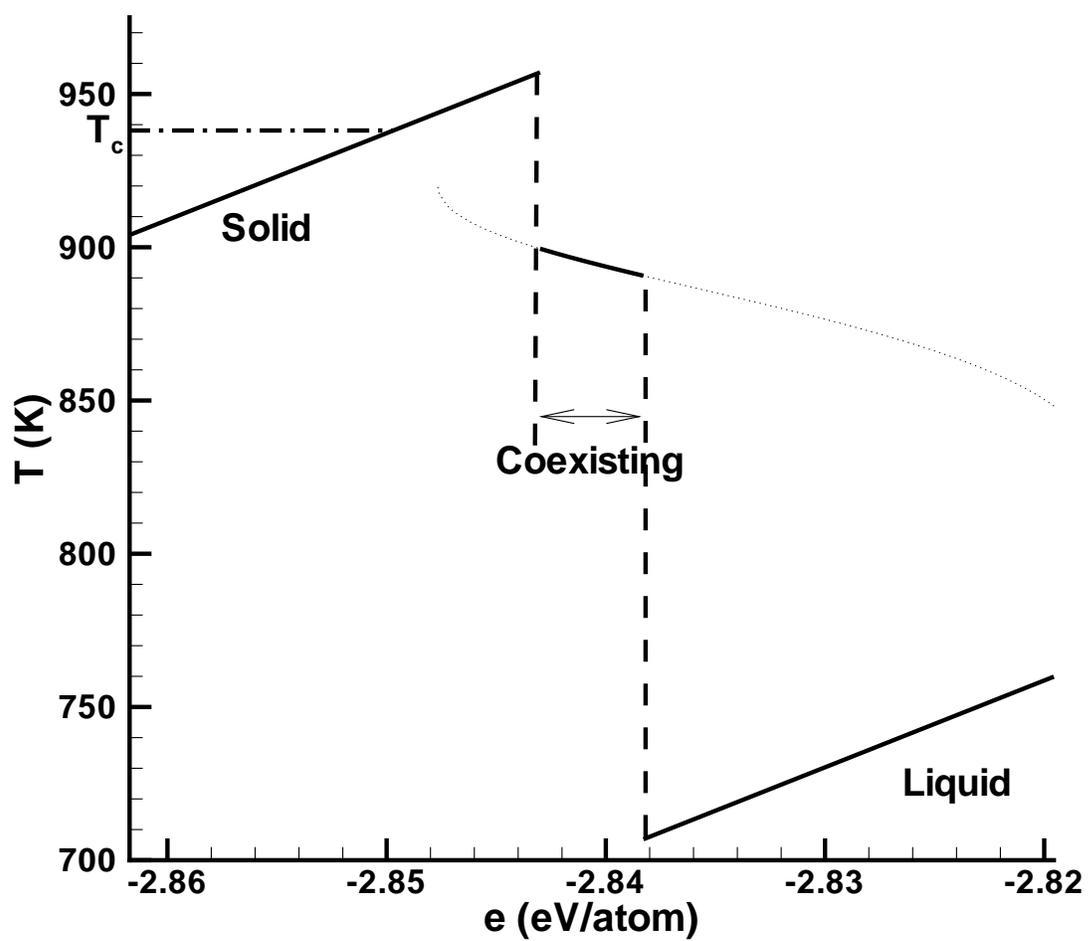}}
\caption{\label{caloric35} Caloric curve for a 3.5 nm Al cluster constructed using MCD theory.}
\end{figure}

In Fig.~\ref{caloric35} we plot the caloric curve constructed by extremizing (\ref{entropy}) for a 3.5 nm Al cluster (which corresponds to 9590 atoms). We have assumed the cluster has only (111) facets and have used material parameters estimated using the glue potential for Al \cite{Ercolessi94,Tosatti95}. In particular these parameters give $\Delta \gamma$ = -2.3 meV$/$\AA$^2$ and $\frac{\Delta\gamma}{\gamma_{sl}} \sim -0.23$ \cite{Tosatti95}, so in a 3.5 nm cluster, $h^* \sim 8 \mbox{\AA} > \xi$. Thus, unless the interaction between the LV and SL interfaces causes $\Delta \gamma$ to change sign, our discussion will be unaffected by the approximation in (\ref{Gamma}). Note the first order character of the transition at the onset of coexistence, consistent with the inequality (\ref{hstar}). In this case we calculate the the onset of coexistence occurs at $h$=19.8 \AA $\,> h^*$. 

We also note the superheating of the solid evident in Fig.~\ref{caloric35}: the cluster does not start to melt until $T > T_c$. In the SM case, where $\Delta \gamma > 0$, a solid cluster begins to melt below $T_c$ because it can convert an increment of surface energy into latent heat. In the NM case, the cluster will not melt until it is favorable to melt a finite volume with $h > h^*$. With the parameter values used here, the temperature of the solid at melting satisfies $T > T_c$. Indeed, it is important to emphasise that the superheated state here is stable rather than metastable. We note however, that for smaller values of $\Delta \gamma$ or $L$, for instance, we have found that the melting temperature can drop below $T_c$. Thus superheating is not a necessary consequence of the model but results from the particular parameters used for the Al (111) surface.  

With these parameters, for $R < 3.4$ nm  we find that the is no energy $E$ where $S_m(h^\dagger,E)>S_s(E)$ and $S_m(h^\dagger,E)>S_l(E)$. In other words the coexisting state is always unstable and the cluster will melt fully without undergoing coexistence when $S_s(E)=S_l(E)$. At this point it is useful to note that the model geometry in Fig.~\ref{geometry} probably overestimates $\Gamma$ for the coexisting particle. Thus we expect the theory here will underestimate the stability of the coexisting state, and overestimate the degree of superheating. In addition, it is important to point out that our model neglects the size-dependence of properties such as the latent heat of melting, L (which appears implicitly in ($\ref{entropy}$) as $S_l^c-S_s^c = L/T_c$).

However we now give an example of this superheating and the corresponding first-order transition to coexistence obtained by MD simulations of Al nanoparticles using the glue potential \cite{Ercolessi94}. With this potential, closed-shell Al clusters with 586 or more atoms favor TO structures \cite{Doye06}. We simulated caloric curves for the 586, 1289, 2406 and 4033-atom TO clusters. At each energy, the cluster was equilibrated for 0.34 ns (with time step $1.35$ fs), then the kinetic energy was averaged over a further 0.34 ns. The energy increment between simulations was 1.0 meV/atom with energies adjusted by a uniform scaling of the kinetic energy. Coexisting states were identified by the appearance of a bimodal distribution of atomic mobilities \cite{Cleveland94, Hendy05b}. 
                                                             
\begin{figure}
\resizebox{\columnwidth}{!}{\includegraphics{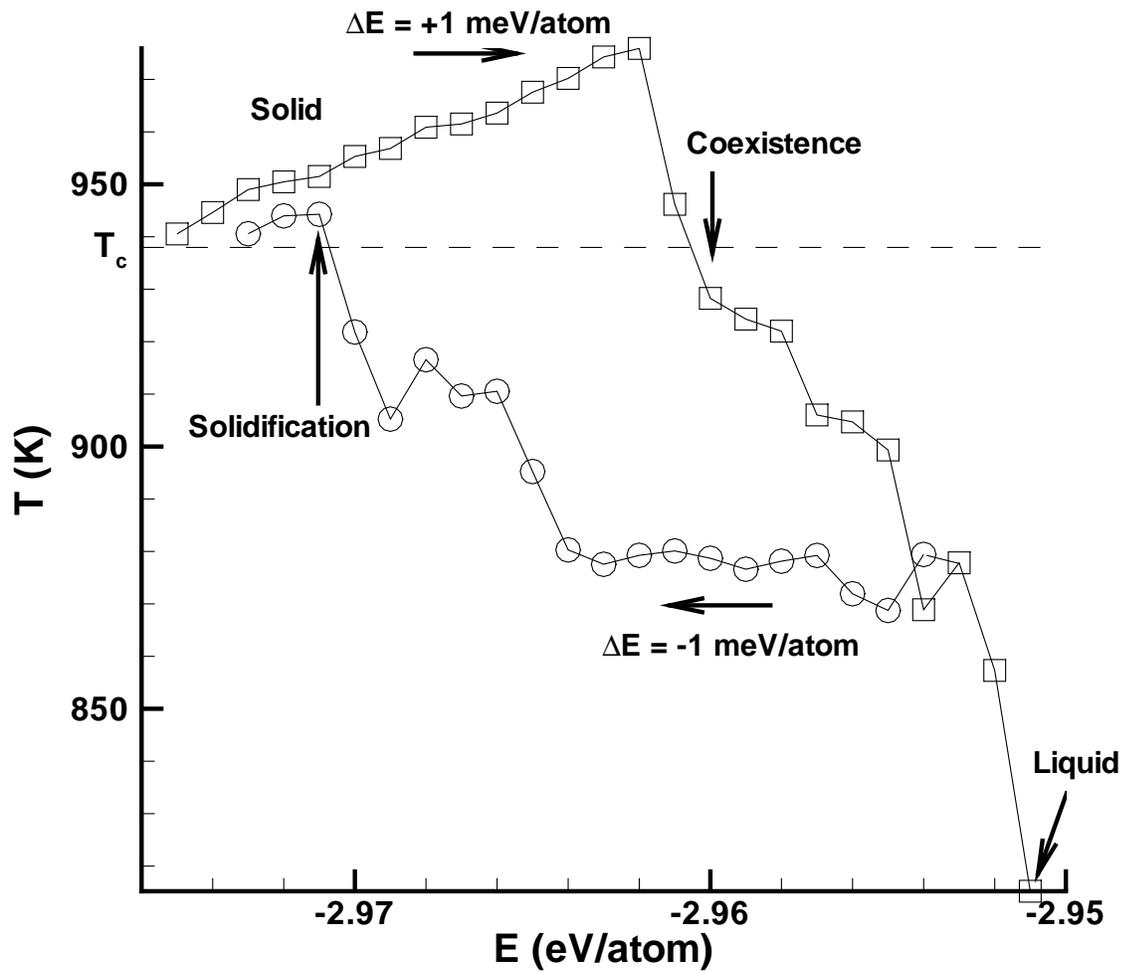}}
\caption{\label{caloric-MD} Caloric curve for a 4033-atom Al cluster. The curve was constructed both by heating from the solid state (squares) and cooling from a coexisting structure (circles) that emerges as the cluster approaches the liquid state.}
\end{figure}

The caloric curve for the 4033 atom cluster is shown in Fig.~\ref{caloric-MD}. Firstly, we note that the transition from the solid to solid-liquid coexistence is first-order, as we expect from ($\ref{hstar}$), with a large drop in temperature at the onset of coexistence. Further, the cluster melting begins at $975$ K which considerably exceeds the bulk melting temperature ($T_c=939$ K using the glue potential). Thus, qualitatively, the 4033-atom cluster shows similar behavior to that predicted by MCD theory. 
 
\begin{figure}
\resizebox{\columnwidth}{!}{\includegraphics{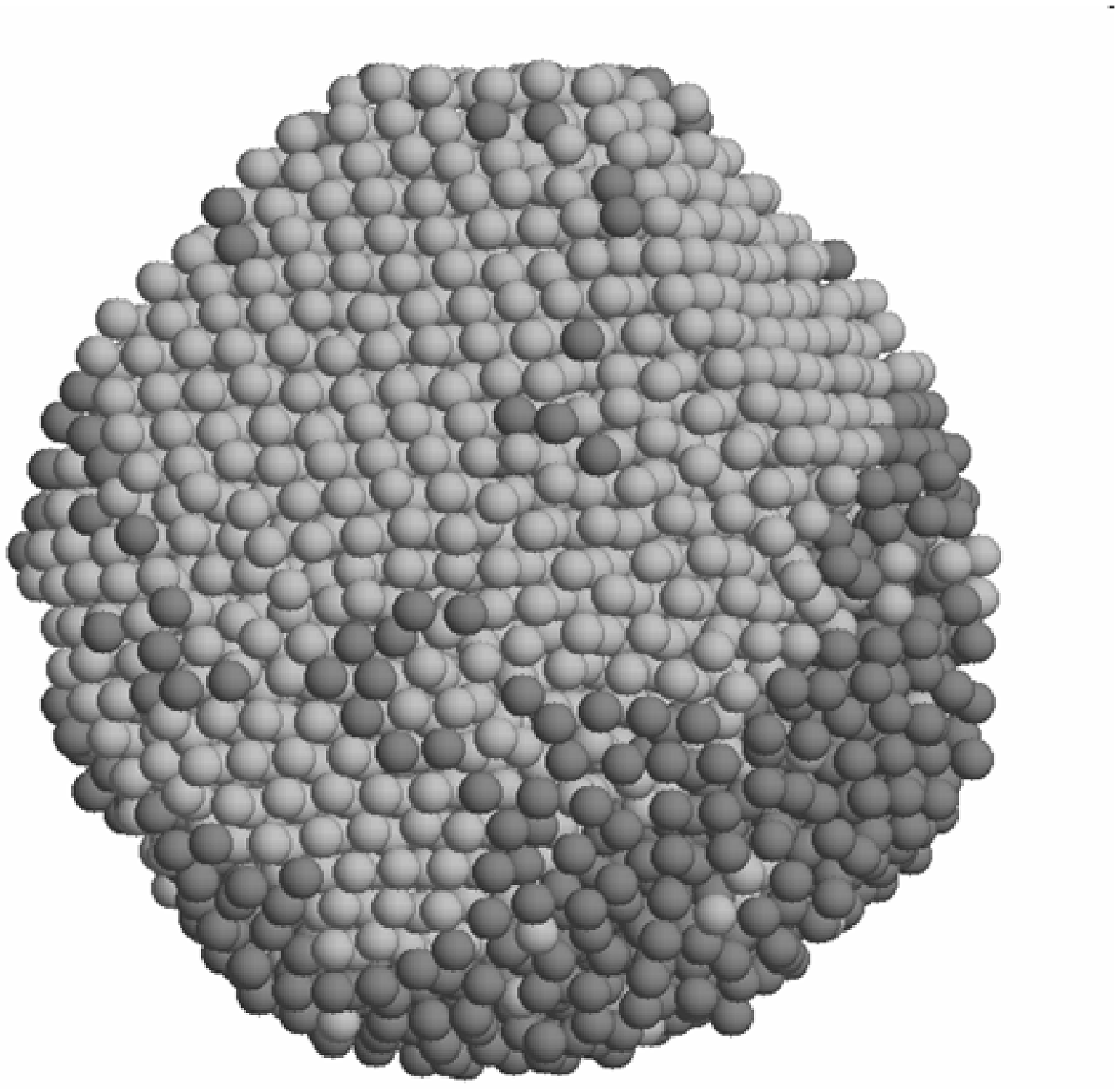}\includegraphics{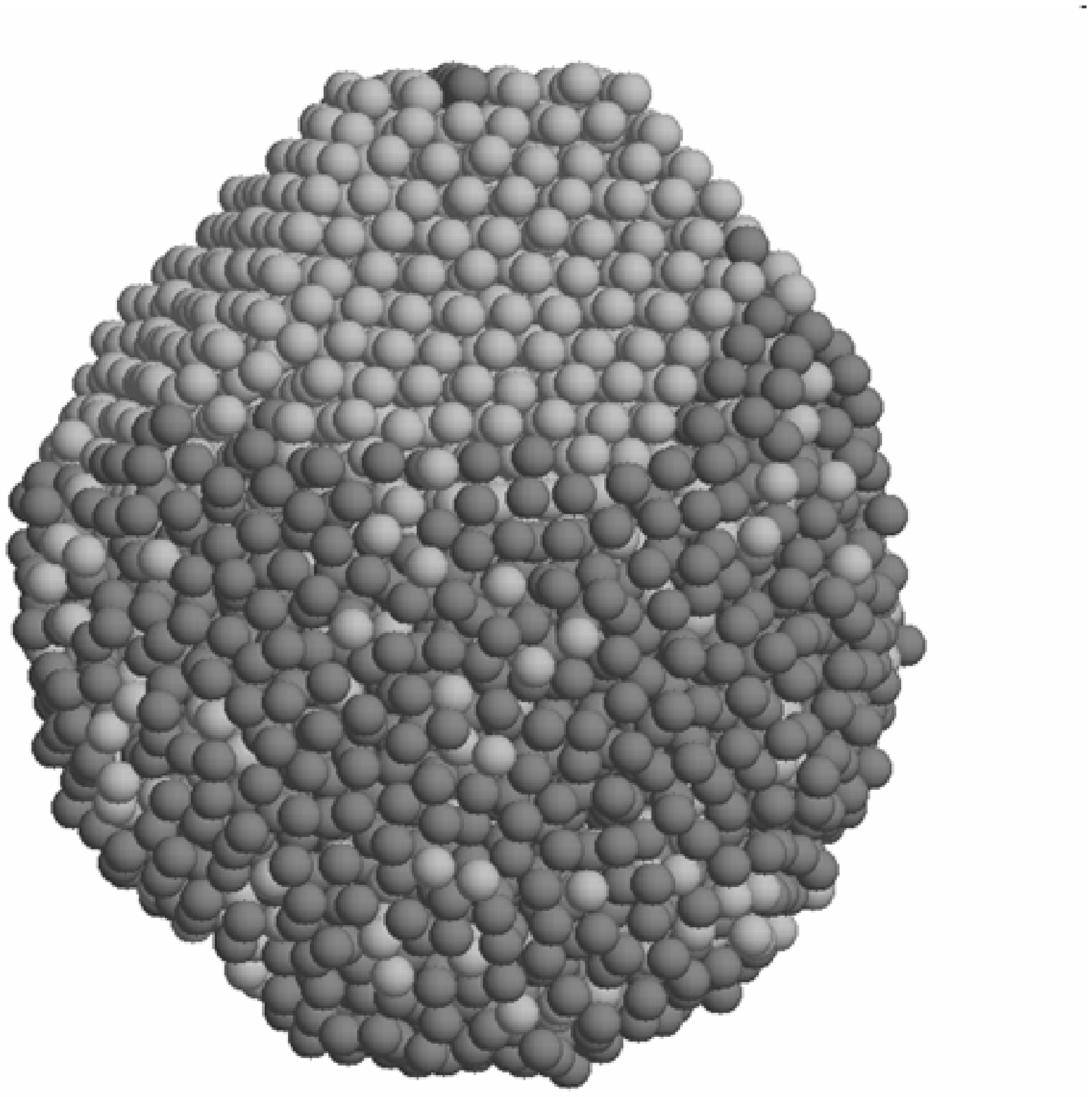}}
\caption{\label{Snaps} Snap shots showing the state of the 4033-atom cluster at $E=-2.966$ eV/atom (left) and $E=-2.960$ eV/atom (right) during the cooling phase of the caloric curve in Fig. \ref{caloric-MD}. Light grey atoms are of low mobility (solid) and the dark grey atoms are of hight mobility (liquid).}
\end{figure}

Of course, the overheated solid could just be metastable rather than stable as suggested by MCD theory. To test this, we selected a cluster state at a point in the coexistence region and removed energy at the corresponding rate. The hysteresis in the cluster state is evident in Fig.~\ref{caloric-MD}, but the cluster fully solidifies at a temperature of 944 K. This shows that the superheated solid is stable relative to the coexisting state over a range of energies. Thus we conclude that the melting temperature of the 4033 cluster lies between 975 and 944 K. Interestingly, the 4033-atom cluster displays two distinct coexisting states shown in Fig.~\ref{Snaps}. In the first state, which is at least metastable, melting is confined largely to two neighboring (100)-facets and the (111)-facet adjoining them. In the second state, all but one of the (100)-facets have melted. 

\begin{figure}
\resizebox{\columnwidth}{!}{\includegraphics{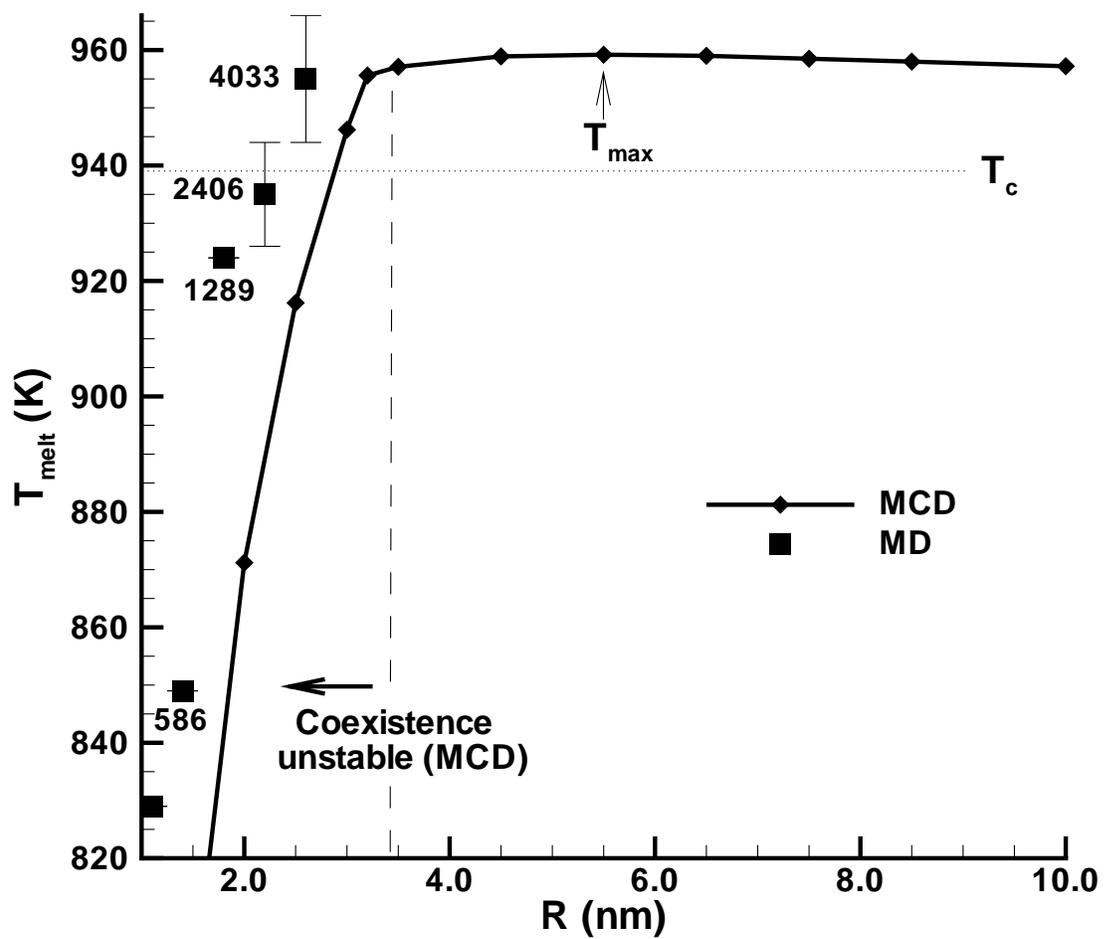}}
\caption{\label{Tmelt} The dependence of melting point on size as calculated using MCD theory for an Al (111)-faceted particle (solid lines), and several data points calculated from MD simulations.}
\end{figure}

Fig.~\ref{Tmelt} compares the melting points predicted by MCD theory with those obtained by MD simulation. Also indicated on the plot is the threshold for coexistence predicted by the theory. In the MD simulations only the 2406 and 4033-atom clusters exhibited static phase coexistence prior to melting. No indications of static, transient or dynamic coexistence were seen prior to melting in the simulations of the 586-atom cluster although the 1289-atom cluster displayed transient coexistence (i.e. dynamic coexistence between the solid and a solid-liquid state, suggesting that the coexisting state is metastable). Thus the threshold for the stability of coexistence lies between the 1289-atom and the 2406-atom clusters. This is a smaller threshold than we obtained from MCD theory, but as noted earlier, our model geometry (Fig.~\ref{geometry}) is likely underestimate the stability of the coexisting cluster. We note that MCD theory predicts that the superheating effect will peak at a cluster radius of $R=5.5$ nm. As $R \rightarrow \infty$, the melting temperature approaches $T_c$.  
           
In previous work, we have constructed microcanonical caloric curves for Ag, Cu, Ni \cite{Hendy05b}, Pd \cite{Hendy06} and Pb \cite{Hendy05a} nanoparticles using molecular dynamics with embedded atom method \cite{Foiles86} potentials. Our initial studies of Pb icosahedra with NM (111)-facets \cite{Pluis87} found that in particles with radii less than 2 nm,  solid-liquid coexistence was absent prior to melting, and that at sizes above this the transition to coexistence was first-order. We did not see superheating in the Pb icosahedra - we recall this is not a necessary consequence of MCD theory. In Ag, Cu, Ni and Pd we saw coexistence in much smaller particles. Furthermore, the transition to coexistence in these clusters was always continuous. In Ag we saw evidence for a coexisting solid-liquid state in a 309-atom cluster, realized as dynamic coexistence between a solid-liquid state and a fully solid state. Similarly in Pd, we saw a coexisting solid-liquid state in a 309-atom cluster, although it was realized as dynamic coexistence between a solid-liquid state and a fully liquid state. Static solid-liquid coexistence was observed in both Ni and Cu 561-atom icosahedra. These previous findings are consistent with our considerations here, where clusters with SM facets exhibit a continuous transition to coexistence down to sizes of $R \sim 1$ nm, and clusters with NM facets avoid coexistence in sizes below $R \sim 2$ and exhibit a first-order transition to coexistence at sizes above this. Thus the stability of phase coexistence in a cluster is strongly dependent on the sign of $\Delta \gamma$.     

In conclusion, we have found theoretical evidence for superheating in clusters with NM surface facets in the microcanonical ensemble. The superheating is associated with a minimum stable liquid volume, which also requires a first-order transition at the onset of melting. Unlike previous observations of a metastable superheated state supported clusters \cite{Metois89}, the microcanonical superheated state discussed here in free clusters is expected to be stable.   



\clearpage
\thispagestyle{empty}

\clearpage
\thispagestyle{empty}

\clearpage
\thispagestyle{empty}

\clearpage
\thispagestyle{empty}

\clearpage
\thispagestyle{empty}


\begin{thebibliography}{27}
\expandafter\ifx\csname natexlab\endcsname\relax\def\natexlab#1{#1}\fi
\expandafter\ifx\csname bibnamefont\endcsname\relax
  \def\bibnamefont#1{#1}\fi
\expandafter\ifx\csname bibfnamefont\endcsname\relax
  \def\bibfnamefont#1{#1}\fi
\expandafter\ifx\csname citenamefont\endcsname\relax
  \def\citenamefont#1{#1}\fi
\expandafter\ifx\csname url\endcsname\relax
  \def\url#1{\texttt{#1}}\fi
\expandafter\ifx\csname urlprefix\endcsname\relax\def\urlprefix{URL }\fi
\providecommand{\bibinfo}[2]{#2}
\providecommand{\eprint}[2][]{\url{#2}}

\bibitem[{\citenamefont{Gross}(2001)}]{Gross01}
\bibinfo{author}{\bibfnamefont{D.H.E.}~\bibnamefont{Gross}},
  \emph{\bibinfo{title}{Microcanonical thermodynamics: phase transitions in finite systems}}
  \bibinfo{journal}{Lecture notes in Physics} \textbf{\bibinfo{volume}{66}},
  \bibinfo{publisher}{World Scientific},(\bibinfo{year}{2001}).

\bibitem[{\citenamefont{Buffat and Borel}(1976)}]{Buffat76}
\bibinfo{author}{\bibfnamefont{Ph.}~\bibnamefont{Buffat}}
\bibnamefont{and} \bibinfo{author}{\bibfnamefont{J-P.}~\bibnamefont{Borel}},
\bibinfo{journal}{Phys. Rev. A} \textbf{\bibinfo{volume}{13}}, 
\bibinfo{pages}{2287} (\bibinfo{year}{1976}).
  
\bibitem[{\citenamefont{Ben David et~al.}(1995)}]{BenDavid95}
\bibinfo{author}{\bibfnamefont{T.} \bibnamefont{Ben David}}  
\bibinfo{author}{\bibfnamefont{Y.} \bibnamefont{Lereah}},
\bibinfo{author}{\bibfnamefont{G.} \bibnamefont{Deutscher}},
\bibinfo{author}{\bibfnamefont{R.} \bibnamefont{Kofman}},
\bibnamefont{and} \bibinfo{author}{\bibfnamefont{P.} \bibnamefont{Cheyssac}},
  \bibinfo{journal}{Phil. Mag. A} \textbf{\bibinfo{volume}{71}},
  \bibinfo{pages}{1135-1143} (\bibinfo{year}{1995}).

\bibitem[{\citenamefont{Baletto and Ferrando}(2005)}]{Baletto05}
\bibinfo{author}{\bibfnamefont{F.}~\bibnamefont{Baletto}}
\bibnamefont{and} \bibinfo{author}{\bibfnamefont{R.}~\bibnamefont{Ferrando}},
\bibinfo{journal}{Rev. Mod. Phys.} \textbf{\bibinfo{volume}{77}}, 
\bibinfo{pages}{371-423} (\bibinfo{year}{2005}).
  
\bibitem[{\citenamefont{Schmidt et~al.}(2001)}]{Haberland01}
\bibinfo{author}{\bibfnamefont{M.} \bibnamefont{Schmidt}},
\bibinfo{author}{\bibfnamefont{R.} \bibnamefont{Kusche}},
\bibinfo{author}{\bibfnamefont{T.} \bibnamefont{Hippler}},
\bibinfo{author}{\bibfnamefont{J.} \bibnamefont{Donges}},
\bibinfo{author}{\bibfnamefont{W.} \bibnamefont{Kronmuller}},
\bibinfo{author}{\bibfnamefont{B.} \bibnamefont{von Issendorff}},
\bibnamefont{and}
\bibinfo{author}{\bibfnamefont{H.} \bibnamefont{Haberland}},
\bibinfo{journal}{Phys. Rev. Lett.} \textbf{\bibinfo{volume}{86}},
  \bibinfo{pages}{1191-1194} (\bibinfo{year}{2001}).

\bibitem[{\citenamefont{Schmidt et~al.}(1998)}]{Haberland98}
\bibinfo{author}{\bibfnamefont{M.} \bibnamefont{Schmidt}},
\bibinfo{author}{\bibfnamefont{R.} \bibnamefont{Kusche}},
\bibinfo{author}{\bibfnamefont{B.} \bibnamefont{von Issendorf}}
\bibnamefont{and}
\bibinfo{author}{\bibfnamefont{H.} \bibnamefont{Haberland}},
\bibinfo{journal}{Nature} \textbf{\bibinfo{volume}{393}},
  \bibinfo{pages}{238-240} (\bibinfo{year}{1998}).

\bibitem[{\citenamefont{Metois and Heyraud}(1989)}]{Metois89}
\bibinfo{author}{\bibfnamefont{J.~J.}~\bibnamefont{Metois}}
\bibnamefont{and} \bibinfo{author}{\bibfnamefont{J.~C.}~\bibnamefont{Heyraud}},
\bibinfo{journal}{J. Phys. (Paris)} \textbf{\bibinfo{volume}{50}}, 
\bibinfo{pages}{3175} (\bibinfo{year}{1989}).
   
\bibitem[{\citenamefont{Breaux et al}(2003)}]{Breaux03}
\bibinfo{author}{\bibfnamefont{G.~A.} \bibnamefont{Breaux}},
\bibinfo{author}{\bibfnamefont{R.~C.} \bibnamefont{Benirschke}},
\bibinfo{author}{\bibfnamefont{T.} \bibnamefont{Sugai}},
\bibinfo{author}{\bibfnamefont{B.~S.} \bibnamefont{Kinnear}}
\bibnamefont{and}
\bibinfo{author}{\bibfnamefont{M.~F.} \bibnamefont{Jarrold}},
\bibinfo{journal}{Phys. Rev. Lett.} \textbf{\bibinfo{volume}{91}},
\bibinfo{pages}{215508} (\bibinfo{year}{2003}).


\bibitem[{\citenamefont{Breaux et al}(2006)}]{Breaux06}
  \bibinfo{author}{\bibfnamefont{S.} \bibnamefont{Krishnamurty}},  
  \bibinfo{author}{\bibfnamefont{S.} \bibnamefont{Chacko}},  
  \bibinfo{author}{\bibfnamefont{D.~G.} \bibnamefont{Kanhere}},
  \bibinfo{author}{\bibfnamefont{G.~A.} \bibnamefont{Breaux}},
  \bibinfo{author}{\bibfnamefont{C.~M.} \bibnamefont{Neal}}
  \bibnamefont{and}
  \bibinfo{author}{\bibfnamefont{M.~F.} \bibnamefont{Jarrold}},
  \bibinfo{journal}{Phys. Rev. B} \textbf{\bibinfo{volume}{73}},
  \bibinfo{pages}{045406} (\bibinfo{year}{2006}).
 
\bibitem[{\citenamefont{Tartaglino et~al}(2005)}]{Tosatti05}
\bibinfo{author}{\bibfnamefont{U.} \bibnamefont{Tartaglino}},
\bibinfo{author}{\bibfnamefont{T.} \bibnamefont{Zykova-Timan}},
\bibinfo{author}{\bibfnamefont{F.} \bibnamefont{Ercolessi}}
\bibnamefont{and} \bibinfo{author}{\bibfnamefont{E.} \bibnamefont{Tosatti}},
\bibinfo{journal}{Phys. Rep.}  \textbf{\bibinfo{volume}{411}},
\bibinfo{pages}{291} (\bibinfo{year}{2005}).

\bibitem[{\citenamefont{van der Veen et al}(1990)}]{vanderVeen90}
\bibinfo{author}{\bibfnamefont{J.~F.} \bibnamefont{van der Veen}},
\bibinfo{author}{\bibfnamefont{B.} \bibnamefont{Pluis}} \bibnamefont{and}
  \bibinfo{author}{\bibfnamefont{A.~W.} \bibnamefont{Denier van der Gon}},
in \emph{\bibinfo{title}{Kinetics of Ordering and Growth at Surfaces}}
\bibinfo{pages}{343-354},  (\bibinfo{publisher}{Plenum Press}, 
\bibinfo{address}{New York}, \bibinfo{year}{1990}). 
  
\bibitem[{\citenamefont{Pluis et al}(1987)}]{Pluis87}
\bibinfo{author}{\bibfnamefont{B.} \bibnamefont{Pluis}},
\bibinfo{author}{\bibfnamefont{A.~W.}~\bibnamefont{Denier van der Gon}},
\bibinfo{author}{\bibfnamefont{J.~W.~M.}~\bibnamefont{Frenken}}
\bibnamefont{and}
\bibinfo{author}{\bibfnamefont{J.~F.}~\bibnamefont{van der Veen}},
\bibinfo{journal}{Phys. Rev. Lett.} \textbf{\bibinfo{volume}{59}},
\bibinfo{pages}{2678-2681} (\bibinfo{year}{1987}).

\bibitem[{\citenamefont{Denier van der Gon et~al}(1990)}]{Denier90}
\bibinfo{author}{\bibfnamefont{A.~W.} \bibnamefont{Denier van der Gon}},
\bibinfo{author}{\bibfnamefont{R.~J.} \bibnamefont{Smith}},
\bibinfo{author}{\bibfnamefont{J.~M.} \bibnamefont{Gay}},
\bibinfo{author}{\bibfnamefont{D.~J.} \bibnamefont{O'Connor}}
\bibnamefont{and} \bibinfo{author}{\bibfnamefont{J.~F.} \bibnamefont{van der Veen}},
\bibinfo{journal}{Surf. Sci.}  \textbf{\bibinfo{volume}{227}},
\bibinfo{pages}{143} (\bibinfo{year}{1990}).

\bibitem[{\citenamefont{Molenbroek and Frenken}(1994)}]{Frenken94}
\bibinfo{author}{\bibfnamefont{A.~M.}~\bibnamefont{Molenbroek}} \bibnamefont{and}
  \bibinfo{author}{\bibfnamefont{J.~W.~M.}~\bibnamefont{Frenken}},
  \bibinfo{journal}{Phys. Rev. B} \textbf{\bibinfo{volume}{50}},
  \bibinfo{pages}{11132} (\bibinfo{year}{1994}).

\bibitem[{\citenamefont{Carnevali et~al}(1987)}]{Carnevali87}
\bibinfo{author}{\bibfnamefont{P.}~\bibnamefont{Carnevali}},
\bibinfo{author}{\bibfnamefont{F.}~\bibnamefont{Ercolessi}}  \bibnamefont{and}
  \bibinfo{author}{\bibfnamefont{E.}~\bibnamefont{Tosatti}},
  \bibinfo{journal}{Phys. Rev. B} \textbf{\bibinfo{volume}{36}},
  \bibinfo{pages}{6701} (\bibinfo{year}{1987}).
   
\bibitem[{\citenamefont{Di Tolla et al}(1995)}]{Tosatti95}
\bibinfo{author}{\bibfnamefont{F.~D.} \bibnamefont{Di Tolla}},
\bibinfo{author}{\bibfnamefont{F.} \bibnamefont{Ercolessi}} \bibnamefont{and}
\bibinfo{author}{\bibfnamefont{E.} \bibnamefont{Tosatti}},
  \bibinfo{journal}{Phys. Rev. Lett.} \textbf{\bibinfo{volume}{74}},
  \bibinfo{pages}{3201-3204} (\bibinfo{year}{1995}). 

\bibitem[{\citenamefont{L\"ummen and Kraska}(2005)}]{Lummen05}
\bibinfo{author}{\bibfnamefont{N.}~\bibnamefont{L\"ummen}} \bibnamefont{and}
  \bibinfo{author}{\bibfnamefont{T.}~\bibnamefont{Kraska}},
  \bibinfo{journal}{Phys. Rev. B} \textbf{\bibinfo{volume}{71}},
  \bibinfo{pages}{205403} (\bibinfo{year}{2005}).

\bibitem[{\citenamefont{Hendy et~al.}(2003)}]{Hendy03}
\bibinfo{author}{\bibfnamefont{S.}~\bibnamefont{Hendy}},
\bibinfo{author}{\bibfnamefont{S.~A.}~\bibnamefont{Brown}} \bibnamefont{and}
  \bibinfo{author}{\bibfnamefont{M.}~\bibnamefont{Hyslop}},
  \bibinfo{journal}{Phys. Rev. B} \textbf{\bibinfo{volume}{68}},
  \bibinfo{pages}{241403(R)} (\bibinfo{year}{2003}).
  
\bibitem[{\citenamefont{Lynden-Bell and Wales}(1994)}]{Lynden-Bell}
  \bibinfo{author}{\bibfnamefont{R.~M.} \bibnamefont{Lynden-Bell}}
  \bibnamefont{and} \bibinfo{author}{\bibfnamefont{D.~J.} \bibnamefont{Wales}},
  \bibinfo{journal}{J. Chem. Phys.} \textbf{\bibinfo{volume}{101}},
  \bibinfo{pages}{1460} (\bibinfo{year}{1994}).

\bibitem[{\citenamefont{Hendy}(2005)}]{Hendy05a}
  \bibinfo{author}{\bibfnamefont{S.~C.} \bibnamefont{Hendy}},
  \bibinfo{journal}{Phys. Rev. B} \textbf{\bibinfo{volume}{71}},
  \bibinfo{pages}{115404} (\bibinfo{year}{2005}).

\bibitem[{\citenamefont{Berry et~al.}(1994)}]{Berry88}
  \bibinfo{author}{\bibfnamefont{R.~S.} \bibnamefont{Berry}},
  \bibinfo{author}{\bibfnamefont{T.~L.} \bibnamefont{Beck}},
  \bibinfo{author}{\bibfnamefont{H.~L.} \bibnamefont{Davis}}
  \bibnamefont{and} 
  \bibinfo{author}{\bibfnamefont{J.} \bibnamefont{Jellinek}},
  \bibinfo{journal}{Adv. Chem. Phys.} \textbf{\bibinfo{volume}{70}},
  \bibinfo{pages}{75} (\bibinfo{year}{1988}).
  
\bibitem[{\citenamefont{Cleveland et~al.}(1994)}]{Cleveland94}
\bibinfo{author}{\bibfnamefont{C.~L.} \bibnamefont{Cleveland}},
\bibinfo{author}{\bibfnamefont{U.} \bibnamefont{Landman}},
\bibnamefont{and}
\bibinfo{author}{\bibfnamefont{W.~D.} \bibnamefont{Luedtke}},
  \bibinfo{journal}{J. Phys. Chem.} \textbf{\bibinfo{volume}{98}},
  \bibinfo{pages}{6272-6279} (\bibinfo{year}{1994}).

\bibitem[{\citenamefont{Schebarchov and Hendy}(2005)}]{Hendy05b}
\bibinfo{author}{\bibfnamefont{D.}~\bibnamefont{Schebarchov}} \bibnamefont{and}
\bibinfo{author}{\bibfnamefont{S.~C.}~\bibnamefont{Hendy}},
  \bibinfo{journal}{J. Chem. Phys.} \textbf{\bibinfo{volume}{123}},
  \bibinfo{pages}{104701} (\bibinfo{year}{2005}).
    
\bibitem[{\citenamefont{Nielsen et~al.}(1994)}]{Nielsen94}
\bibinfo{author}{\bibfnamefont{O.~H.} \bibnamefont{Nielsen}},
\bibinfo{author}{\bibfnamefont{J.~P.} \bibnamefont{Sethna}},
\bibinfo{author}{\bibfnamefont{P.} \bibnamefont{Stoltze}},
\bibinfo{author}{\bibfnamefont{K.~W.} \bibnamefont{Jacobsen}},
\bibnamefont{and}
  \bibinfo{author}{\bibfnamefont{J.~K.} \bibnamefont{Norskov}},
  \bibinfo{journal}{Europhys. Lett.} \textbf{\bibinfo{volume}{26}},
  \bibinfo{pages}{51-56} (\bibinfo{year}{1994}).

\bibitem[{\citenamefont{Ercolessi and Adams}(1994)}]{Ercolessi94}
\bibinfo{author}{\bibfnamefont{F.}~\bibnamefont{Ercolessi}},
\bibnamefont{and} \bibinfo{author}{\bibfnamefont{J.~B.}~\bibnamefont{Adams}},
\bibinfo{journal}{Europhys. Lett.} \textbf{\bibinfo{volume}{26}},
\bibinfo{pages}{583} (\bibinfo{year}{1994}).

\bibitem[{\citenamefont{Noya et al}(2006)}]{Doye06}
\bibinfo{author}{\bibfnamefont{E.~G.} \bibnamefont{Noya}},
\bibinfo{author}{\bibfnamefont{J.~P.~K.} \bibnamefont{Doye}} \bibnamefont{and}
  \bibinfo{author}{\bibfnamefont{F.} \bibnamefont{Calvo}},
  \bibinfo{journal}{Phys. Rev. B} \textbf{\bibinfo{volume}{73}},
  \bibinfo{pages}{125407} (\bibinfo{year}{2006})

\bibitem[{\citenamefont{Schebarchov and Hendy}(2005)}]{Hendy06}
\bibinfo{author}{\bibfnamefont{D.}~\bibnamefont{Schebarchov}} \bibnamefont{and}
\bibinfo{author}{\bibfnamefont{S.~C.}~\bibnamefont{Hendy}},
  \bibinfo{journal}{Phys. Rev. B} \textbf{\bibinfo{volume}{73}},
  \bibinfo{pages}{121402(R)} (\bibinfo{year}{2006}).

\bibitem[{\citenamefont{Foiles et al}(1986)}]{Foiles86}
\bibinfo{author}{\bibfnamefont{S.~M.} \bibnamefont{Foiles}},
\bibinfo{author}{\bibfnamefont{M.~I.}~\bibnamefont{Baskes}}
\bibnamefont{and}
\bibinfo{author}{\bibfnamefont{M.~S.}~\bibnamefont{Daw}},
\bibinfo{journal}{Phys. Rev. B} \textbf{\bibinfo{volume}{33}},
\bibinfo{pages}{7983-7991} (\bibinfo{year}{1986}).

\end{thebibliography}
\end{document}